\documentclass[12pt,fleqn]{article}    
\usepackage[latin1]{inputenc}    
\usepackage[intlimits]{amsmath}
\usepackage{amssymb}
\usepackage{graphicx}
\usepackage{epsfig}
\begin{document}
\baselineskip 18pt
\title{Relativistic Covariance and Quark-Diquark Wave Functions for Baryons
\thanks{Supported by the Forschungszentrum FZ J\"ulich (COSY)}}
\author{M. Dillig
\thanks{mdillig@theorie3.physik.uni-erlangen.de}                    \\ 
Institute for Theoretical Physics III \thanks{preprint FAU-TP3-06/Nr. 05}\\
 University of Erlangen-N\"urnberg\\
 Staudtstr. 7, D-91058 Erlangen, Germany}
\date{}
\baselineskip 18pt
\begin{titlepage}

\maketitle

\begin{abstract}
We derive covariant wave functions for hadrons composed of two
constituents for arbitrary Lorentz
boosts. Focussing explicitly on baryons as quark-diquark systems, we reduce their
manifestly covariant Bethe-Salpeter equation to covariant
3-dimensional forms by projecting on the relative quark-diquark
energy. Guided by a phenomenological multi gluon exchange
representation of covariant confining kernels, we derive explicit solutions for harmonic confinement and
for the MIT Bag Model. We briefly sketch implications of breaking the spherical symmetry of the ground state and the
transition from the instant form to the light cone via the 
infinite momentum frame. 
\end{abstract}

\end{titlepage}
%
%
%
%
\newpage
\setcounter{page}{2}
Modern electron and hadron accelerators investigate the internal structure and dynamics of hadrons and implications for the non perturbative regime of QCD: detailed 
information is extracted from scattering
experiments at large momentum transfers of typically 1 GeV and
beyond. The corresponding form factors map out the various internal
(generalized) charge distributions and provide stringent information
on the underlying quark and gluon degrees of freedom. Presently
various experiments are ongoing with electron and proton beams at various labs (1,2). 
\vskip 0.2cm
With energy and momentum transfers increasing bébeyond a typical scale of
1 GeV, at least approximate covariance is an important ingredient in any
microscopic parametrization; however, its quantitative
importance and implementation is still under investigations for practical 
calculations and quantitatively
not fully understood. Traditionally
performed in the instant (equal time) form,
observables, such as form factors, are sensitive to the
pertinent problem of center-of-mass (CM) corrections for the many-body
problem and from substantial effects from Lorentz contraction at
increasing momentum transfers. Both for the CM corrections (3-6) and the formulation of covariant hadronic wave functions, various
recipes have been developed and applied in practical calculations
(7-28). A
possible alternative, the evaluation of form factors on the light cone,
where Lorentz boosts are kinematical (29), has so far entered
only selectively in applications at low
and intermediate scattering energies;
beyond that, such an approach suffers from other deceases, such as the
loss of strict rotational invariance (30). As in general the
construction of  boosted, Lorentz contracted wave functions in the instant form 
is nearly as
complicated as the solution of the full problem (as here the boost are dynamical and depend explicitly on the interaction kernel employed in the problem), in most practical
applications simple kinematical prescriptions for the
rescaling of the coordinate along the direction of the momentum
transfer are applied (examples are given ref. (31,32)); however, specific
questions, as the dependence of Lorentz corrections on the confining
kernel  in quark models, are not addressed.
\vskip 0.2cm
In this note we formulate an economical model for hadron baryon
wave functions, which leads to results suitable for practical
applications. As it is our main goal to end up with analytical formulae,
we model the baryon - in the following we focus on the proton, though
our approach is fairly general - as a quark-diquark system (33) and restrict
ourselves, for technical simplicity and 
without any loss of generality, to spin-isospin scalar
diquarks; including axial diquarks or modelling mesons as $q \overline q$
systems basically only introduces a more angular momentum structure of the
corresponding amplitudes (see our comment below).
\vskip 0.2cm
Our starting point is the manifestly covariant 4-dimensional
Bethe-Salpeter equation (34) 
\begin{equation}
\Gamma = K \, G\, \Gamma \quad \mbox{and} \quad \Psi = G \, \Gamma
\end{equation}
with the vertex function and the Bethe-Salpeter amplitude $ \Gamma $
and $\Psi $, respectively, and the interaction kernel K. In the
two-body Greens function for the quark and diquark masses
m and m*,  we fix the relative energy dependence from the covariant
projection on the diquark (35)
\begin{eqnarray}
G (P, q) = \frac{q\!\!\!/ + m}{\nu (q^{2} - m ^{2})+
 (1-\nu)((Q-q)^{2} - m ^{2}) - i \epsilon} \; i \pi (1-\nu)\\
\delta _{+} 
\left (\nu(q^2-m^2)-(1-\nu) ((Q - q)^{2} - m^{^2})) \right ).
\end{eqnarray}
As we would like to connect below with the boost of the MIT bag,
which we simulate for an static (infinite) spectator-diquark mass,
we pursue among the infinite many projections the limit $\nu \rightarrow \infty$ 
which  results up to $0 \left ( \frac{q^{2}}{2M} \right ) $ in the
single particle Dirac equation for the quark for systems with
arbitrary  overall 4-momentum $Q = ( E(P) = \sqrt{P^{2} + M^{2}}\, ,
0, 0, P ) $  \\
\begin{eqnarray}
 ( ( \frac{M^2}{2E(P)}+ \frac{P}{E(P)}\, q_{z})-({\bf \alpha q} + \beta m))
\phi(P, {\bf q}) \, =
 \, \frac{1}{E(P)} \int K (p, {\bf q},{\bf k}) \phi (P, {\bf k}) d {\bf k}
\end{eqnarray}\\
with M being the mass of the baryon (the external momentum P is chosen along the z axis).
\vskip 0.2cm
Without any details we add a brief comment on the CM corrections in
our model: evidently there is a direct coupling between the internal
and external momenta {\bf q} and {\bf P}, or equivalently,
between boosts and the CM motion. In the rest
system the leading center-of-mass corrections are absorbed for $
\epsilon = m + \epsilon _{b}$, where $ \epsilon _{b} $ is the binding
energy of the quark in 
\begin{equation}
\left ( \frac{q^{2}}{2 \mu} + \epsilon _{b} - V_{n} (r) \right )
\varphi ({\bf r}) = 0 
\end{equation}
with the reduced mass $1 /\mu \cong 1 /m + 1 /(m + m^{*}) $ for an
arbitrary quark potential $ V_{n} (r)$. 
\vskip 0.2cm
The important step for a practical model is the formulation of a
covariant  interaction kernel in eq.~(4). As the dynamics of the quark
- quark interactions, particularly the microscopic nature of the
confinement, lacks an understanding on the fundamental level of QCD,
all current models in practical calculations rely on phenomenological
formulations of the interaction kernel. Here we
proceed here along similar lines: we parametrize the interaction
kernel as a superposition of appropriately weighted
gluon exchange contributions; quantitative parameters can be
extracted in comparison with studies to baryon spectroscopy, decay
rates or form factors (36). 
Thus we start from the general kernel
\begin{equation}
K(P, q,k) = \sum _{n} \; \frac{k_{n} (P)}{((q - k)^{2} - \mu ^{2} - i
\epsilon )^{n+1}} 
\end{equation}
for arbitrary powers of n (which reflect different parametrizations of
the confining kernel; for n=1 eq.(5) contains the linear confinement in the
Cornell potential (37) and as confirmed from lattice calculations). Upon projecting out the relative energy
dependence this yields the covariant, 3-dimensional kernel 
\begin{equation}
K_{n} (P, q, k) \propto \;_{ {\lim \atop \mu \to 0}} \;
\left(\frac{d}{d \mu^{2}} \right )^n \; \frac{1}{\lambda ^{2} (P) q
_{z} ^{2} + {\bf q}^{2}_{\bot} + \mu ^{2} - i \epsilon} 
\end{equation}
with the "quenching parameter"
\begin{equation}
\lambda (P) = M / \sqrt{M^{2} + P^{2} } \, = M / E(P)
\end{equation}
where we introduced the mass scale $\mu $ (to regularize the Fourier
transform to coordinate space). Already simple power counting signals,
that a kernel with the power $n \geq 1$ leads to confinement with $\sim
r^{2n-1}$. Upon performing the corresponding Fourier transform to
coordinate space and taking the limit  
$ \mu \to 0 $  we find 
\begin{equation}
K_{n} (P,q) \to  (1+\beta)/2\,  V_{n} (\sqrt{(z/\lambda(P))^{2} +
\mbox{\boldmath$\rho$} ^{2})} 
\end{equation}
(with $\bf \rho$ as the perpenticular component to z),where we introduced for convenience 
the particular Dirac structure of the kernel to
facilitate the evaluation of the resulting Dirac equation (more general
Dirac invariants modify in addition the structure of the small
Dirac component; compare our remark below). 
Eliminating the small component in eq. (3) with the kernel from
eq. (6) and upon dropping CM corrections and $ \epsilon ^{2}_{b} $
terms for compactness, we end up with the Schroedinger type equation
for the large component of the Dirac equation 
\begin{eqnarray}
& & \ (2m\epsilon_{b} -\lambda^{2}( q_{z}- \frac{m}{M} P)^2 ) - {\bf{q}
 ^{2}_{\bot}} -  
 \nonumber\\
&  & \qquad \qquad -\, V_{n} \left ( \sqrt{(z /\lambda (P))^{2} + 
\mbox{\boldmath$\rho$} ^{2} ) } / R  \right ) \; u (z, \bf{\rho} ) = 0
\end{eqnarray}
(with the typical length scale R; in the rest system the equation
above reduces to the standard spherical Schr\"odinger type equation for
a particle with mass m). The final equation defines with its
connection to the small component by a simple differentiation the full
relativistic covariant quark - diquark wave function for arbitrary
Lorentz systems. In the equation above we see the shortcoming from the
phenomenological nature of the interaction kernel: we absorb the
explicit $\epsilon$ and P dependence of the kernel in the definition
of the energy scale $V_{n} $ for the confining force; including an
explicit P dependence in $ V_{n}$ would require a detailed knowledge
of its microscopic origin. 
\vskip 0.2cm
Approximate or numerical solutions for eq. (9) can be obtained for
different confining scenarios. Here we discuss briefly two examples,
which allow a rigorous analytical solution:
i. e. harmonic confinement and bag models (in the limit   
$ n \to \infty $ in eq. (6)).
\vskip 0.2cm

\begin{itemize}
\item[-] {\bf Harmonic confinement}: \\
With the harmonic kernel defined as (39)
\begin{eqnarray}
K(P,q,k) & = & - 12/\pi _{{\lim \atop  (\mu \to 0)}} \left ( ( d/d
\mu^{2})^{2} (\mu / 2 + (d/d \mu ^{2}) \mu^{3} /3 \right )
\frac {1}{q^2- \mu ^2 - i \epsilon} \nonumber\\ 
& \to & - \left ( (1/\lambda (P))^{2} (d/d q_{z})^{2} + (d/d{\bf
q}^{2}_{\bot}) \right ) \delta (q_{z}-k_{z}) \delta ({\bf q}_{\bot} -
{\bf k}_{\bot}) \, , 
\end{eqnarray}

the solutions for arbitrary excitations of the baryon are
easily obtained in momentum space. After a redefinition of the
longitudinal momentum
and upon separating the longitudinal and the perpendicular component,
the general solution is given by a product of confluent hypergeometric
functions (40). Here we focus only on the nucleon as the quark -
diquark ground state and obtain explicitly  
\begin{equation}
u(q_{z}, {\bf q}_{\bot}) = N \; e^{- \frac{a^{2}}{2} \;\left ( \lambda
^{2} (q_{z} - \frac{m}{M} P \right ) ^{2} + {\bf q}_{\bot} )^{2}} 
\end{equation}
with the oscillator parameter $ a^{2} = \frac{2}{\sqrt{V_{c}}} $ and
$ V_{c}\propto 1/R^{4} $ 
being the confinement strength.
As expected the standard solution for the spherical harmonic
oscillator is recovered in the rest system, i.e. for P=0 and $\lambda
$(0)=1. As the characteristic result we find a quenching of the
effective P-dependent size parameter 
\begin{equation}
a ^{2}(P)  =  (\lambda (P)a )^{2} = \frac{M^{2}}{P^{2} + M^{2}} \; a^{2} 
\end{equation}
which leads to Lorentz quenching in coordinate space along the z -
axis and thus to a significant increase of the longitudinal
momentum components with increasing P (Fig.1(a,b)); 
\vskip 0.2cm

\item[-] {\bf Bag Model}:\\
As mentioned above we generate the Bag from the transition $ n \to
\infty $ in the power of the gluon-exchange kernel. As we are unable to
present an analytical solution for arbitrary n (a closed solution for
the z-component exists only in the limit of vanishing binding  (40)),
 we first perform the limit $ n \to \infty
$ and then solve the equation 
\begin{equation}
\left (\epsilon _{b} + \frac{P}{M} \; q _{z} + m - ( \alpha \, {\bf q}
+ \beta m) \right ) \; u(z, \mbox{\boldmath$\rho$}) = 0 
\end{equation}
with the boundary condition for the large and small
components at $ z = \lambda (P) R$ for the bag radius R. For the large
component the ground state solution can be represented as 
\begin{equation}
u (z, {\bf \rho }) =
 N \cos (\frac{k_{z}}{\lambda}\, z) \,\, J_{0} ({\bf k_{\bot} \rho })
\end{equation}
where the $\lambda$ dependence of the z-component again reflects the
quenching of the bag. The corresponding boundary condition
 $z=\lambda R$ for $ \mbox{\boldmath$\rho$} = 0$
 for the deformed bag, which reduces to the standard boundary condition for
the spherical bag, 
 can be solved only numerically (41). Characteristic results for
3 different boost momenta are presented for the large and small
component of the bag ground state solution in Fig. 2. 

\vskip 0.2cm
Comparing our findings with simple recipes we
find that a general and simple extension of the parametrization of the
spherical wave functions and momentum distributions in the rest system
to a boosted system, by rescaling the size parameter of the system,
but keeping otherwise the spherical character of the solutions, is
certainly very unsatisfactory and breaks down completely for boost
momenta of typically  $P/M \ge 1$.  Only for very small boost momenta
P simple approximations, such as 
%
\renewcommand{\theequation}{16a}
\begin{equation}
u (z, \mbox{\boldmath$\rho$} , a) \cong u \left ( r, a/ \left (
\sqrt{3} \; \lambda (P) \right) \right) \quad \mbox{and} 
\end{equation}
%
%
\renewcommand{\theequation}{16b}
\begin{equation}
u( z, \mbox{\boldmath$\rho$} , R) \cong \exp ( - (r / (\sqrt{3} \; \lambda
(P) R)) ^{2} \, u (r, R) 
\end{equation}
\renewcommand{\theequation}{\arabic{equation}}
\setcounter{equation}{16}
simulate very qualitatively Lorentz quenching of slowly moving systems.
\end{itemize}
\vskip 0.2cm

The breaking of the spherical symmetry
for moving systems leads to significant modifications of the 
structure of composite objects with increasing boost momenta.
Without entering into details we list just a few consequences.
\vskip 0.2cm
One novel feature for moving particles is the modification of
 their spin-orbit interaction for interaction kernels involving 
different Dirac invariants.
As an example, for a purely scalar confining potential
 the elimination of the lower
component in the Dirac equation leads to
\begin{equation}
V_{ls} \sim (\sigma_z q_z\; + c {\bf \sigma q }) \, V(z, {\bf \rho }) \,
(\sigma_z q_z\, + c^{'} {\bf \sigma q })
\end{equation}
which gives rise to terms $\sim {\bf l_{\perp}s_{\perp} }$ and thus 
to a spin-orbit interaction even for baryons initially in a relative s-state .
\vskip 0.2cm
Similarly, boosting a spherical kernel leads even for baryons with quarks in 
relative s-states - due to the breaking of  spherical symmetry - to
the admixture of additional angular momenta, which significantly enhance
the momentum spectrum of the ground state with increasing q due to their radial structure (in an harmonic basis, dropping normalization constants)

\begin{equation}
\phi(P,{\bf r}) \, \sim \frac{1}{a a^{1/2}(P)}\, 
\exp(-\frac {r^2}{2 a^2}(1+\frac{a^2-a^2(P)}{a^2(P)} \frac{4 \pi}{3} 
Y^2_{10}({\bf \hat r}))),
\end{equation}
which yields for the spectrum in momentum space from the corresponding Fourier transform 
\begin{equation}
\phi(P,{\bf q}) \sim \int\limits_{0}^{1}  dt \, \frac{2 p^2(t)- (qt)^2}{p^5(t)}
\exp(-\frac{(qt)^2}{4 p^2(t)})   
\end{equation}
with $p^2(t) \, = \, (1+\frac{a^2-a^2(P)}{a^2(P)} t^2)/(2 a^2)$.\\

In Fig. 3 a characteristic result for the admixture of higher 
angular momenta to the L=0 ground state of a baryon (with all quarks in relative s states),
 is shown for
different boosts upon projecting out the L=2 orbital angular momentum: evidently, higher angular
 momenta dominate form
factors of hadrons with increasing momentum transfers (for an explicit inclusion of a d-state in the nucleon the present quark-scalar diquark representation has to to include in addition axial diquarks with S = 1, which allow the coupling of the orbital angular momentum L = 2 to the total spin 1/2 of the nucleon).
\vskip 0.2cm
A test for the consistency of the simple approach above is the
transition to the infinite momentum frame for boost momenta
$P \rightarrow \infty$: as in this limit the physics becomes effectively
(1+1)-dimensional, our approach has to merge into the
corresponding dynamical equations on the light cone. Without entering
into details of such a limiting process, we, for simplicity,  only
 sketch here
our reasoning for two spin less bosons with equal masses. With a projection
of the BSE for the two particles symmetrically off their mass shell
(following Blankenbecler and Sugar (35);
an similar result is obtained in the Gross limit
 (42), we obtain 
with the momentum fraction x and the  total longitudinal light cone
 momentum $P_{+}$, from
\begin{eqnarray}
G_{IF} \sim \delta_{+}(q_{0}- \frac{P}{E(P)} q_{z}), \; \;
G_{LC} \sim \delta(q_{-}+ M^{2} x); \, x=\frac{q_{+}}{P_{+}}- \frac{1}{2},
\end{eqnarray}
for $P \rightarrow \infty $ the equation
\begin{equation}
(\frac {M^2}{4}-m^2-y^2-{\bf q_{\perp}}) \Phi (y, {\bf q_{\perp}},M^2)=
\int K(y-y^{'}, {\bf q_{\perp}-q_{\perp}^{'}}),
\Phi (y^{'}, {\bf q_{\perp}^{'}},M^2)\, dy^{'} {\bf  dq_{\perp}^{'}}\\
\end{equation}
with
\begin{equation}
y \equiv y_{ \, IF} \, =\, M \, \frac {q_{z}}{P} , \; \; 
y \equiv y_{\, LC} \, = \, Mx,
\end{equation}
 respectively. Similarly,
the kernels also become identical in the IF and on the
LC in the limit $M \rightarrow \infty$.\\

Summarizing the main findings in this note, we derive
covariant wave functions and their transformation properties in an
analytical quark - diquark model for the baryon. As expected, we find
characteristic modifications from the baryon rest system to moving
Lorentz systems for different confining kernels already for moderate
boost momenta.  
\vskip 0.2cm

Our findings suggest possible extensions and  basic shortcomings of
the model. Evidently its application to mesons as
$q\overline{q}$  systems, towards a more realistic quark-diquark
description of baryons or to genuine   3-quark systems (together with
a systematic inclusion of CM corrections) imposes only technical
problems and is clearly feasible. Here we only mention that the
quenching factor from eq.(7) is recovered in leading order for all
different projections of the BS equation. 
\vskip 0.2cm
A more serious problem is the detailed formulation of confining kernels
 with a finite power in
the inter quark distance r, in particular of the Lorentz
quenching for the kernels itself (in Bag models the dependence is
absorbed in the boundary condition). Here the main problem it the poor phenomenological understanding the confining mechanism: it is
not clear how the full P dependence enters into the kernel 
(for a simple extension compare ref. (43)).
 We feel that a more realistic
formulation of present phenomenological quark models requires
a deeper analytical understanding of confinement, beyond the numerical results from lattice calculations. Here
significant
progress in various directions has been achieved recently, to mention
only the modelling of
confinement of QCD in the Coulomb gauge (44) or the  extension of
concept of instantons to merons as solutions of the classical QCD
equations (45). Given such a more detailed understanding may 
eventually provide some answer to our question at the beginning, i. e.
on the importance of covariance
and the appropriate incorporation of boosts in exclusive processes
at typical energy and momentum transfers of 1 GeV. Valuating
both the approximations and the practicability of calculations
on the one side, and recent progress in light cone physics, such
as restoring rotational invariance (29,46), on the other
side, we expect that 
systematic studies of both electromagnetic and hadronic
exclusive two-body reactions will finally map out
 the
most adequate parametrization of the underlying short
ranged physics of exclusive processes at the low GeV scale, advocating
either
the instant form or the
front form.

\vskip 1.0cm

\begin{figure}
\begin{center}
\includegraphics[width=12.0cm,height=10.0cm,angle=0]{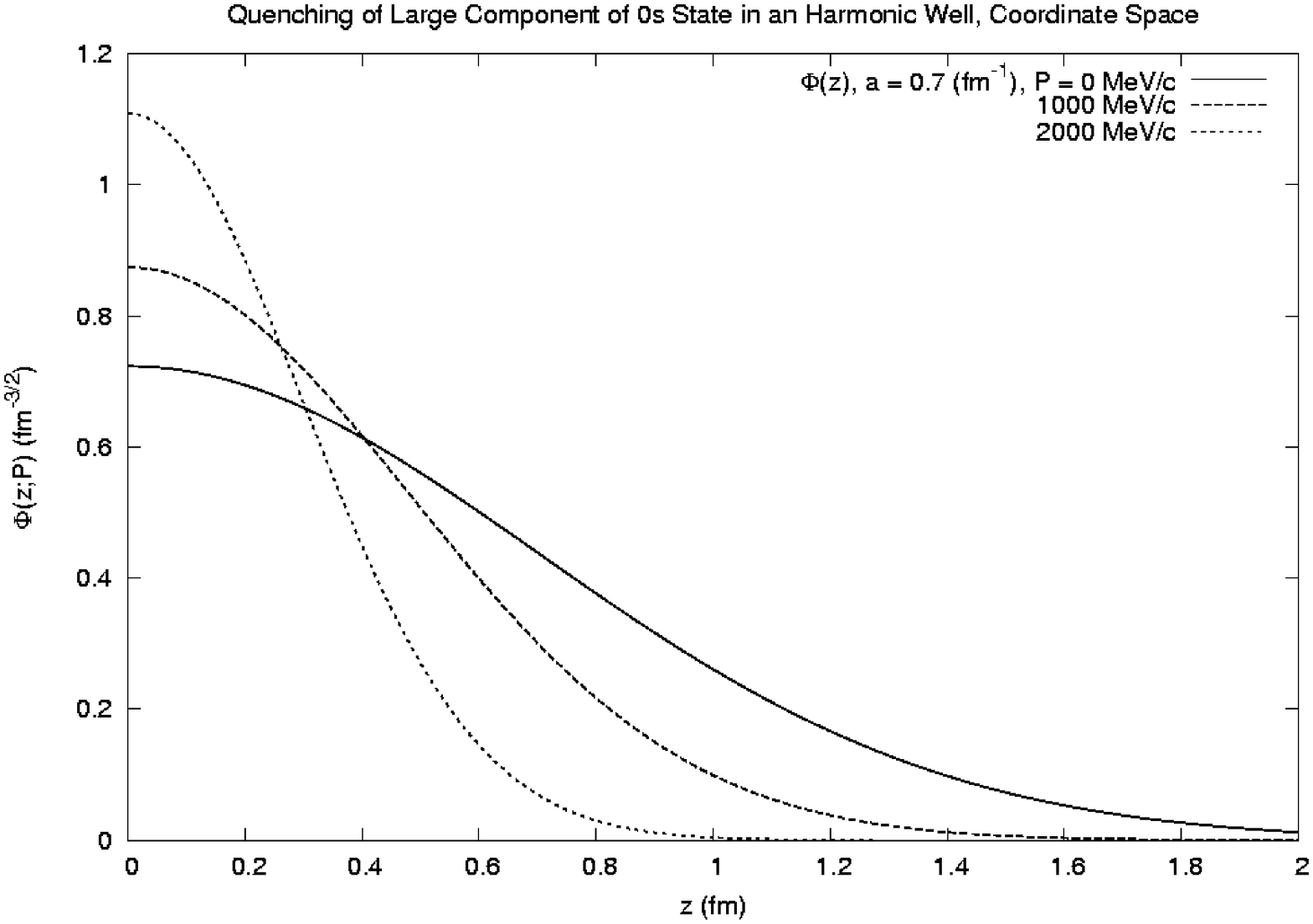}
\includegraphics[width=12.0cm,height=10.0cm,angle=0]{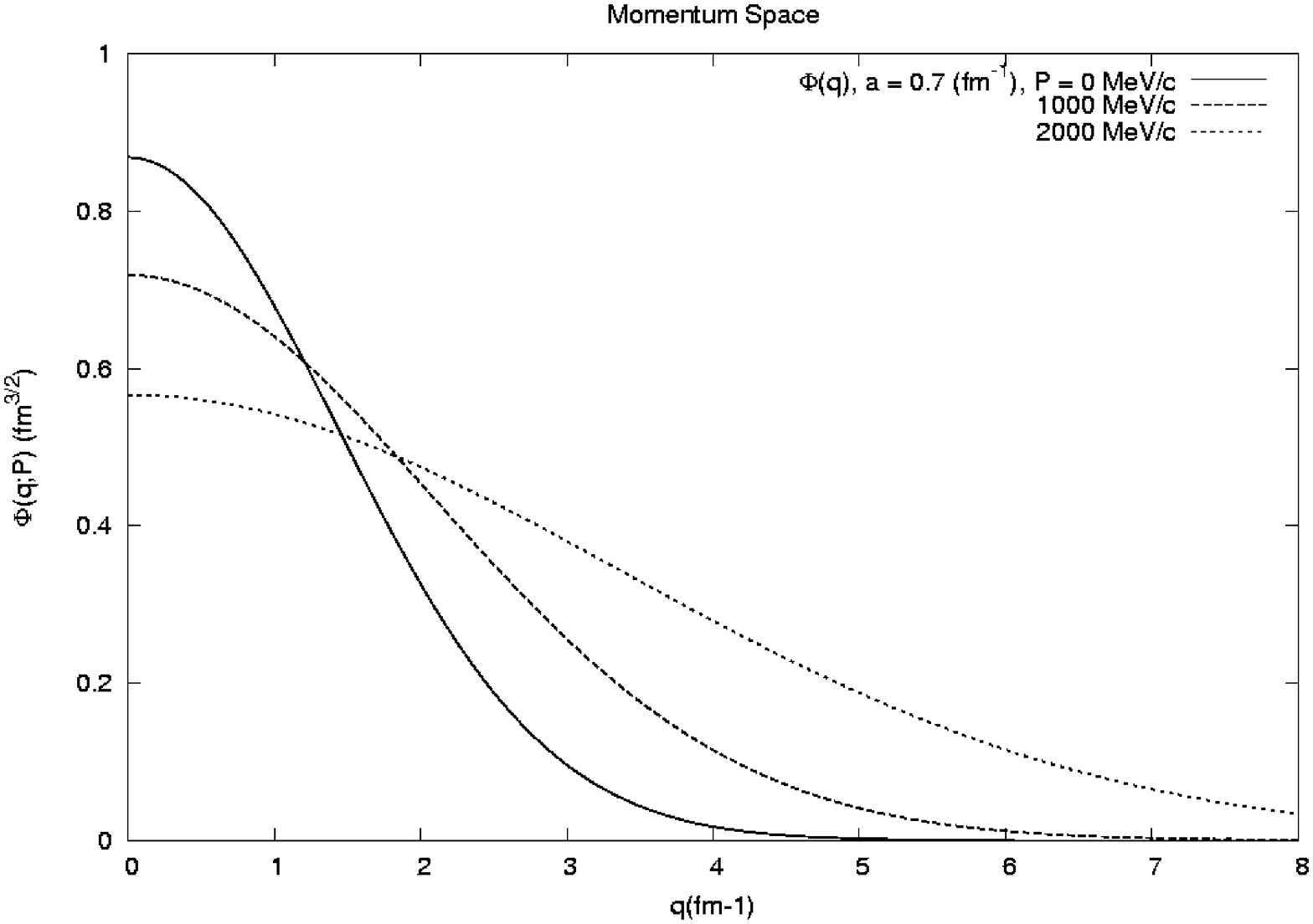}
\caption{Longitudinal dependence of the large component of the harmonic oscillator ground state
in coordinate (a) and momentum space (b) for different boost momenta P.}       
\end{center}
\end{figure}

\begin{figure}
\begin{center}
\includegraphics[width=12.0cm,height=10.0cm,angle=0]{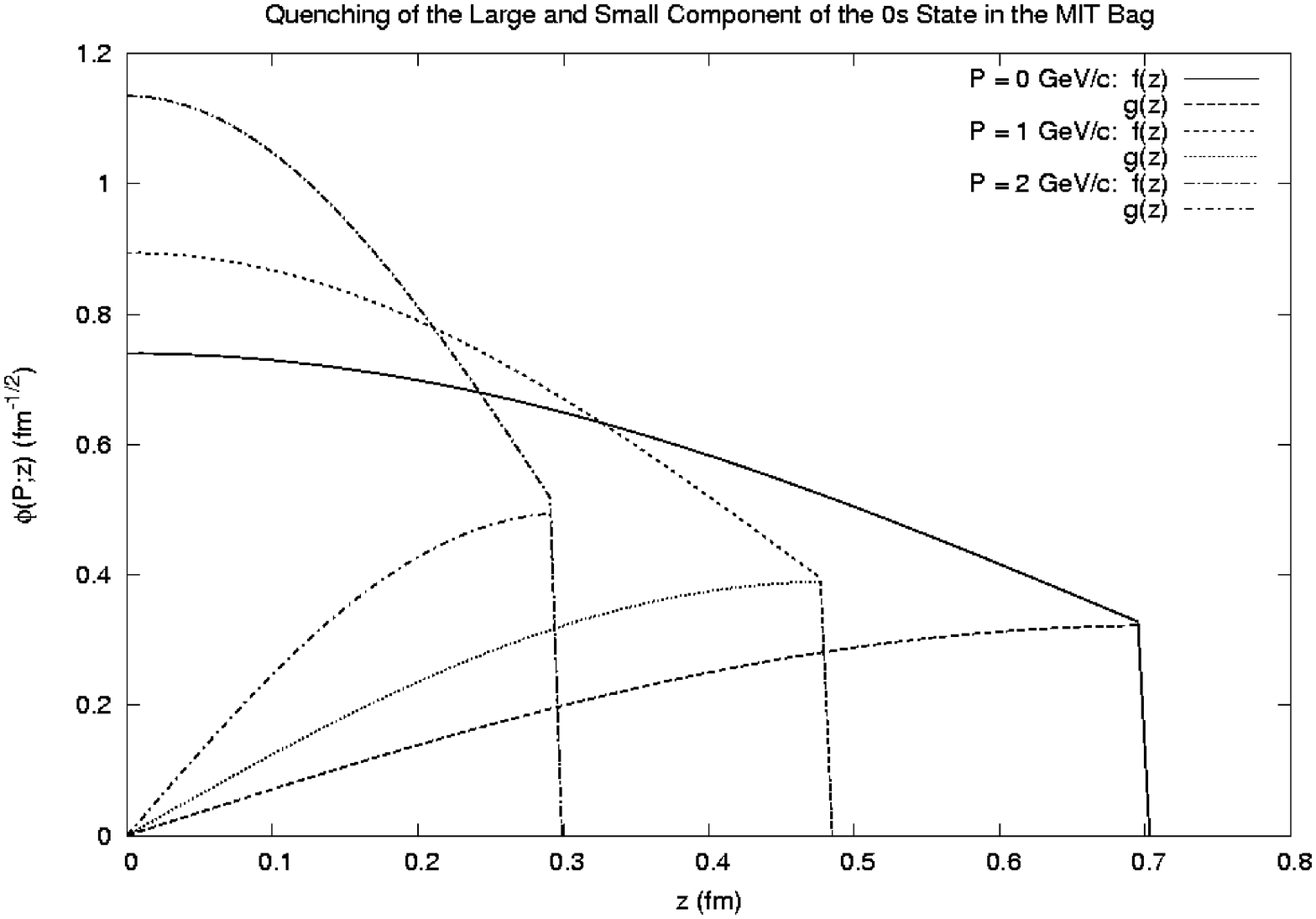}
\caption{
Quenching and boundary conditions for the bag along the boost momenta
P=0, 1 and 2 GeV/c.
 The functions f(z) and g(z) denote the large and small
Dirac components of the ground state wave function for a (static) bag radius 
R = 0.7 fm.}
\end{center}
\end{figure}

\begin{figure}
\begin{center}
\includegraphics[width=12.0cm,height=10.0cm,angle=0]{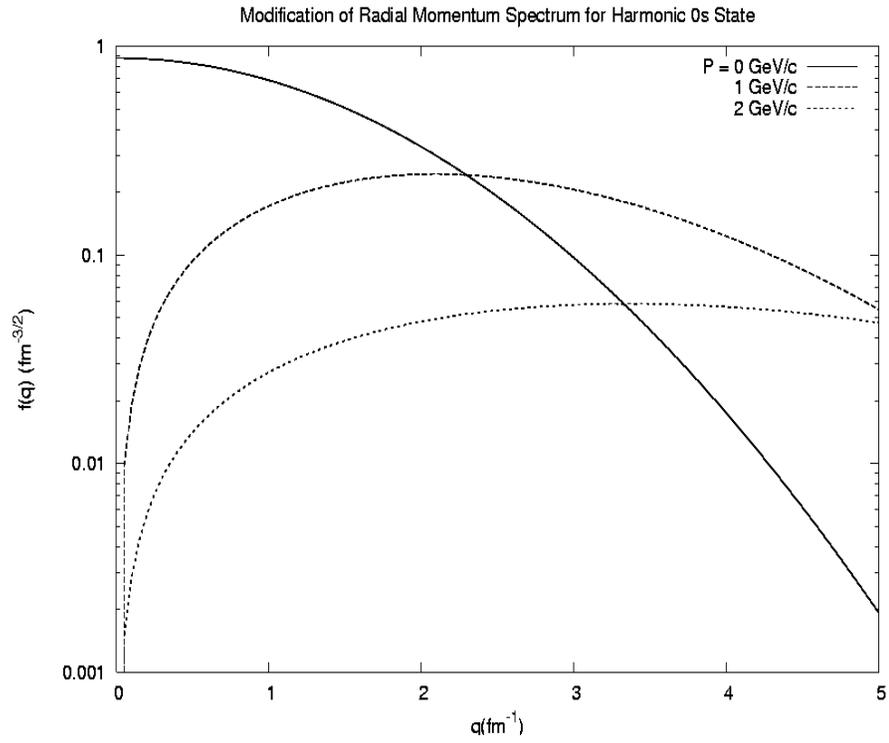}
\caption{D-state admixture to a boosted harmonic oscillator ground state. Compared
are the (spherical L=0) s-wave momentum distribution for P=0 with the L=2 component for P = 1 and 2 GeV/c.} 
\end{center}
\end{figure}

\end{document}